\documentclass[12pt]{iopart}

\usepackage{graphicx}
\usepackage[colorlinks=true]{hyperref}
\usepackage{amssymb}
\usepackage{longtable}
\usepackage{rotating}
\usepackage{color}
\usepackage{epsfig}
\usepackage{units}
\usepackage{wasysym}
\usepackage{subfigure}
\usepackage{parskip}
\usepackage{float}
\begin{document}

\title[Characterization of the Virgo Seismic Environment]{Characterization of the Virgo Seismic Environment}

\author{T.~Accadia$^{11}$, 
F.~Acernese$^{5ac}$, 
P.~Astone$^{8a}$, 
G.~Ballardin$^{2}$, 
F.~Barone$^{5ac}$, 
M.~Barsuglia$^{1}$, 
A.~Basti$^{7ab}$, 
Th.~S.~Bauer$^{14a}$, 
M.~Bebronne$^{11}$, 
M.G.~Beker$^{14a}$
% ...
}
\address{$^{1}$Laboratoire AstroParticule et Cosmologie (APC) Universit\'e Paris Diderot, CNRS: IN2P3, CEA: DSM/IRFU, Observatoire de Paris, 10 rue A.Domon et L.Duquet, 75013 Paris - France}
\address{$^{2}$European Gravitational Observatory (EGO), I-56021 Cascina (PI), Italy}
\address{$^{3}$INFN, Sezione di Firenze, I-50019 Sesto Fiorentino$^a$; Universit\`a degli Studi di Urbino 'Carlo Bo', I-61029 Urbino$^b$, Italy}
\address{$^{4}$INFN, Sezione di Genova;  I-16146  Genova, Italy}
\address{$^{5}$INFN, Sezione di Napoli $^a$; Universit\`a di Napoli 'Federico II'$^b$ Complesso Universitario di Monte S.Angelo, I-80126 Napoli; Universit\`a di Salerno, Fisciano, I-84084 Salerno$^c$, Italy}
\address{$^{6}$INFN, Sezione di Perugia$^a$; Universit\`a di Perugia$^b$, I-06123 Perugia,Italy}
\address{$^{7}$INFN, Sezione di Pisa$^a$; Universit\`a di Pisa$^b$; I-56127 Pisa; Universit\`a di Siena, I-53100 Siena$^c$, Italy}
\address{$^{8}$INFN, Sezione di Roma$^a$; Universit\`a 'La Sapienza'$^b$, I-00185 Roma, Italy}
\address{$^{9}$INFN, Sezione di Roma Tor Vergata$^a$; Universit\`a di Roma Tor Vergata, I-00133 Roma$^b$; Universit\`a dell'Aquila, I-67100 L'Aquila$^c$, Italy}
\address{$^{10}$LAL, Universit\'e Paris-Sud, IN2P3/CNRS, F-91898 Orsay$^a$; ESPCI, CNRS,  F-75005 Paris$^b$, France}
\address{$^{11}$Laboratoire d'Annecy-le-Vieux de Physique des Particules (LAPP), Universit\'e de Savoie, CNRS/IN2P3, F-74941 Annecy-Le-Vieux, France}
\address{$^{12}$Laboratoire Kastler Brossel, ENS, CNRS, UPMC, Universit\'e Pierre et Marie Curie, 4 Place Jussieu, F-75005 Paris, France}
\address{$^{13}$Laboratoire des Mat\'eriaux Avanc\'es (LMA), IN2P3/CNRS, F-69622 Villeurbanne, Lyon, France}
\address{$^{14}$Nikhef, Science Park, Amsterdam, the Netherlands$^a$; VU University Amsterdam, De Boelelaan 1081, 1081 HV Amsterdam, the Netherlands$^b$}
\address{$^{15}$Universit\'e Nice-Sophia-Antipolis, CNRS, Observatoire de la C\^ote d'Azur, F-06304 Nice$^a$; Institut de Physique de Rennes, CNRS, Universit\'e de Rennes 1, 35042 Rennes$^b$, France}
\address{$^{16}$INFN, Gruppo Collegato di Trento$^a$ and Universit\`a di Trento$^b$,  I-38050 Povo, Trento, Italy;   INFN, Sezione di Padova$^c$ and Universit\`a di Padova$^d$, I-35131 Padova, Italy}
\address{$^{17}$IM-PAN 00-956 Warsaw$^a$; Astronomical Observatory Warsaw University 00-478 Warsaw$^b$; CAMK-PAN 00-716 Warsaw$^c$; Bia{\l}ystok University 15-424 Bia{\l}ystok$^d$; IPJ 05-400 \'Swierk-Otwock$^e$; Institute of Astronomy 65-265 Zielona G\'ora$^f$,  Poland}
\address{$^{18}$RMKI, H-1121 Budapest, Konkoly Thege Mikl\'os \'ut 29-33, Hungary}
\address{$^{19}$Physics and Astronomy, Carleton College, Northfield, MN, 55057, USA}
\ead{coughlim@carleton.edu}

\begin{abstract}
The Virgo gravitational wave detector is an interferometer (ITF) with 3km arms located in Pisa, Italy. From July to October 2010, Virgo performed its third science run (VSR3) in coincidence with the LIGO detectors. Despite several techniques adopted to isolate the interferometer from the environment, seismic noise remains an important issue for Virgo. Vibrations produced by the detector infrastructure (such as air conditioning units, water chillers/heaters, pumps) are found to affect Virgo's sensitivity, with the main coupling mechanisms being through beam jitter and scattered light processes. The Advanced Virgo (AdV) design seeks to reduce ITF couplings to environmental noise by having most vibration-sensitive components suspended and in-vacuum, as well as muffle and relocate loud machines. During the months of June and July 2010, a Guralp-3TD seismometer was stationed at various locations around the Virgo site hosting major infrastructure machines. Seismic data were examined using spectral and coherence analysis with seismic probes close to the detector. The primary aim of this study was to identify noisy machines which seismically affect the ITF environment and thus require mitigation attention. Analyzed machines are located at various distances from the experimental halls, ranging from 10m to 100m. An attempt is made to measure the attenuation of emitted noise at the ITF and correlate it to the distance from the source and to seismic attenuation models in soil.
\end{abstract}

\maketitle

\section{Introduction}

The general theory of relativity predicts that all accelerating objects with non-symmetric mass distributions produce gravitational waves (GW) \footnote{As presented at the Gravitational-wave Physics and Astronomy Workshop in Milwaukee, Wisconsin, January 26-29, 2011.}. LIGO (the Laser Interferometer Gravitational-Wave Observatory) \cite{LIGO} and Virgo \cite{VIRGO} experiments seek to directly detect GWs and use them to study astrophysical sources. 

Virgo, located in Cascina, Italy, consists of a laser Michelson interferometer with 3km long, Fabry Perot resonant optical cavities in its arms. With respect to other similar detectors, Virgo has enhanced sensitivity between 10 and 100Hz due to the seismic isolation performance of the ``super-attenuators'' to which the test masses are suspended \cite{Superattenuator}. Beginning in 2011, the detector expects to undergo upgrades, known as Advanced Virgo (AdV), to improve its sensitivity by one order of magnitude \cite{AdVirgo}.

\begin{figure*}[hbtp!]
 \centering \subfigure{}
 \includegraphics[width=3in]{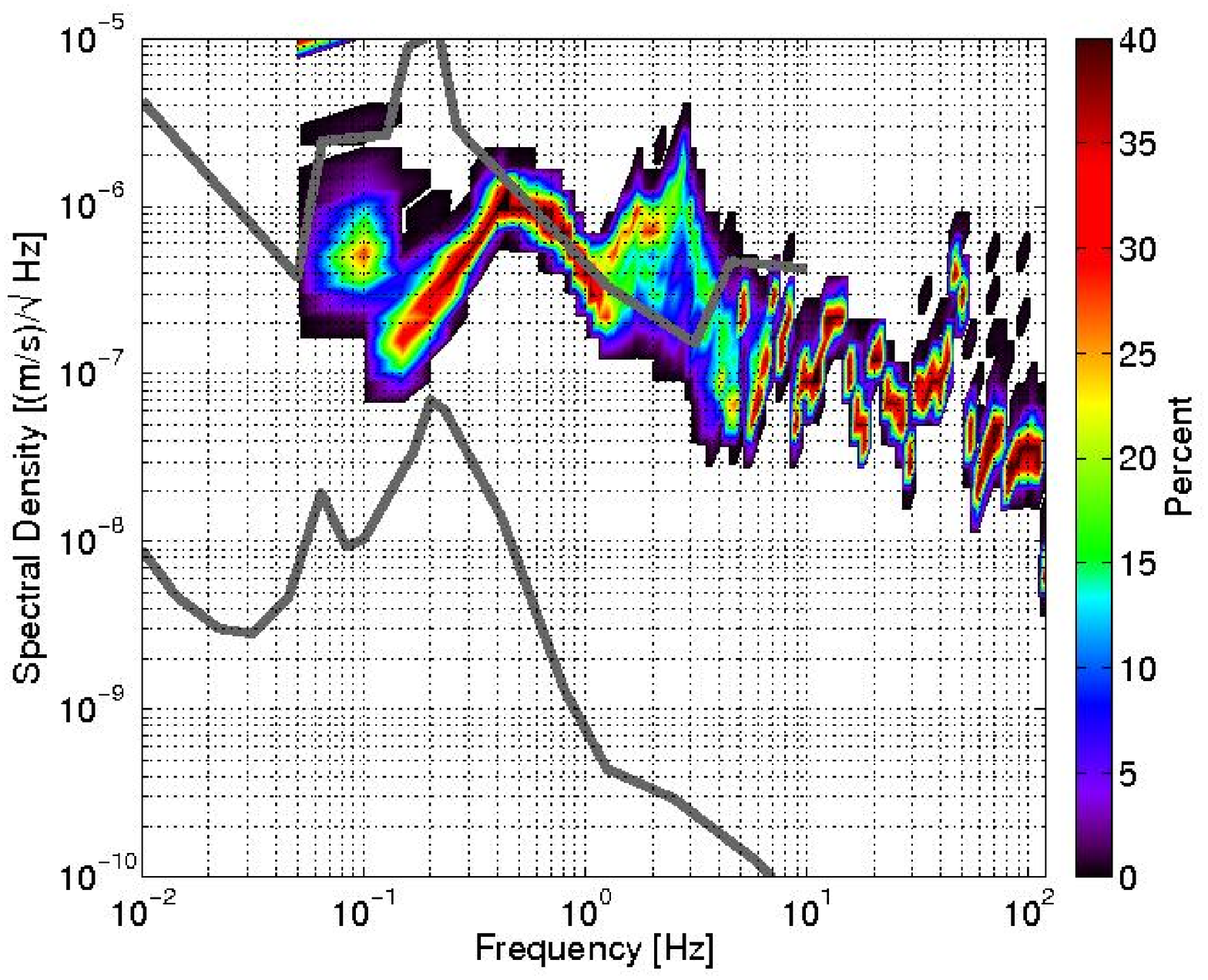}
 \centering \subfigure{}
 \includegraphics[width=3in]{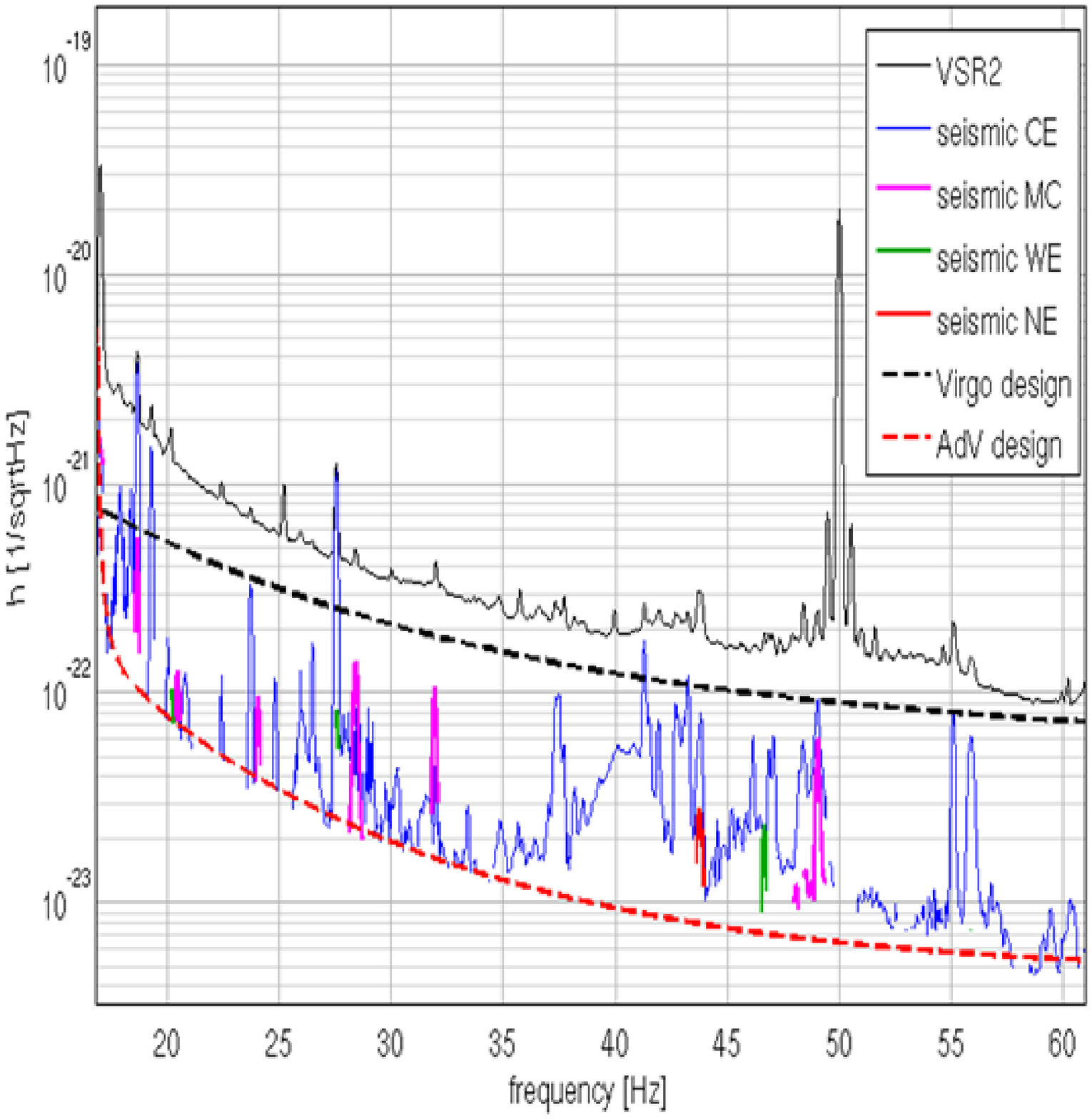}
 \caption{Left: Variation in the root spectral density of the Guralp seismometer at the Virgo site during a day in June of 2010. The two solid curves correspond to the Peterson low- and high-noise models \cite{Peterson}. Right: Plot of the VSR2 sensitivity as well as the coherence projection of seismic probes on the sensitivity.}
 \label{fig:Noisy}
\end{figure*}

Seismic noise places a limit on Virgo's detection sensitivity. In particular, the main source of seismic noise in the 10 to 100Hz band at the Virgo site is the detector's infrastructure machines (please see the left of Fig.~\ref{fig:Noisy} for a comparison with a low noise model). Although Virgo's mirrors are well-isolated from local seismic activity by suspension systems made of multi-stage pendulums, seismic noise remains a concern. The residual ground motion causes ``diffused light'' \cite{DiffusedLight}.  Because of unavoidable imperfections in the detector's optical components, some tiny fraction of light can exit the main optical path and strike a surface that is connected to the ground, and thus be subject to the local seismic field. When this light scatters off of objects connected to the ground (such as optical components on external tables used for detector controls), it is often diffused over a wide solid angle. A fraction of the light can re-enter the main beam path, but with a noisy phase modulated by the seismically excited scattering object. This phenomenon functions as additional noise in the GW channel and limits its sensitivity.

Because seismic events couple in this way into the GW channel, it is necessary to understand and attenuate the noise from the local seismic environment. Many of the strong lines seen during VSR2 can be attributed to seismic sources (please see the right of Fig.~\ref{fig:Noisy}). To assist in the identification effort, the Virgo detector is supplemented with several types of environmental sensors, including seismometers and accelerometers, that monitor the local environment \cite{VirgoEnvironment}. These channels are used to detect environmental disturbances that can couple to the GW channel and are placed in sensitive areas of the interferometer. To reduce the influence of anthropogenic noise, during the Advanced Virgo upgrade, machines that are identified as seismically and acoustically affecting the interferometer will be replaced, moved, or isolated \cite{AdVirgo}. The current proposal is to move all chillers, water pumps, air compressors, and air conditioners from their current locations to other areas further from the interferometer. By placing these machines on their own isolated platforms and by adopting techniques to reduce noise emission and propagation, Virgo hopes to significantly reduce their effects on the detector. To prioritize the actions required to reduce noise at Virgo, a careful documentation of the present local seismic environment is necessary, as well as identification among the infrastructure machines of relevant sources of seismic noise reaching the detector's sensitive components. The work described hereafter is part of this effort.

During the months of June and July 2010, a Guralp-3TD seismometer was stationed to monitor floor vibrations at various technical areas around the Virgo site which host several infrastructure machines (i.e. water chillers, heaters, pumps). The power spectra of the measured seismic data were examined, and several single sources were identified using correlation of the seismic data with synchronous records of the infrastructure machines working status (i.e. temperature monitors, voltage sniffers). Then, seismic data were correlated with the synchronous record of permanent seismic probes close to the detector. Studying the calculated coherence and power spectra, we selected ``relevant'' sources whose noise emissions reach the detector and measure the noise attenuation between the sources and detector. We also measured approximate noise reduction factors as a function of the distance between the machine and the detector.

Section 2 discusses the methods used when taking and analyzing data. Section 3 describes examples of noise emissions from the sources found in this analysis at the various locations, while Section 4 provides an estimate for the attenuation of the noise spectral peaks as a function of distance. Section 5 presents our conclusions.

\section{Characterization Methods}

\subsection{Measurement Setup}

\begin{figure*}[hbtp!]
 \centering \subfigure{}
 \includegraphics[width=3in]{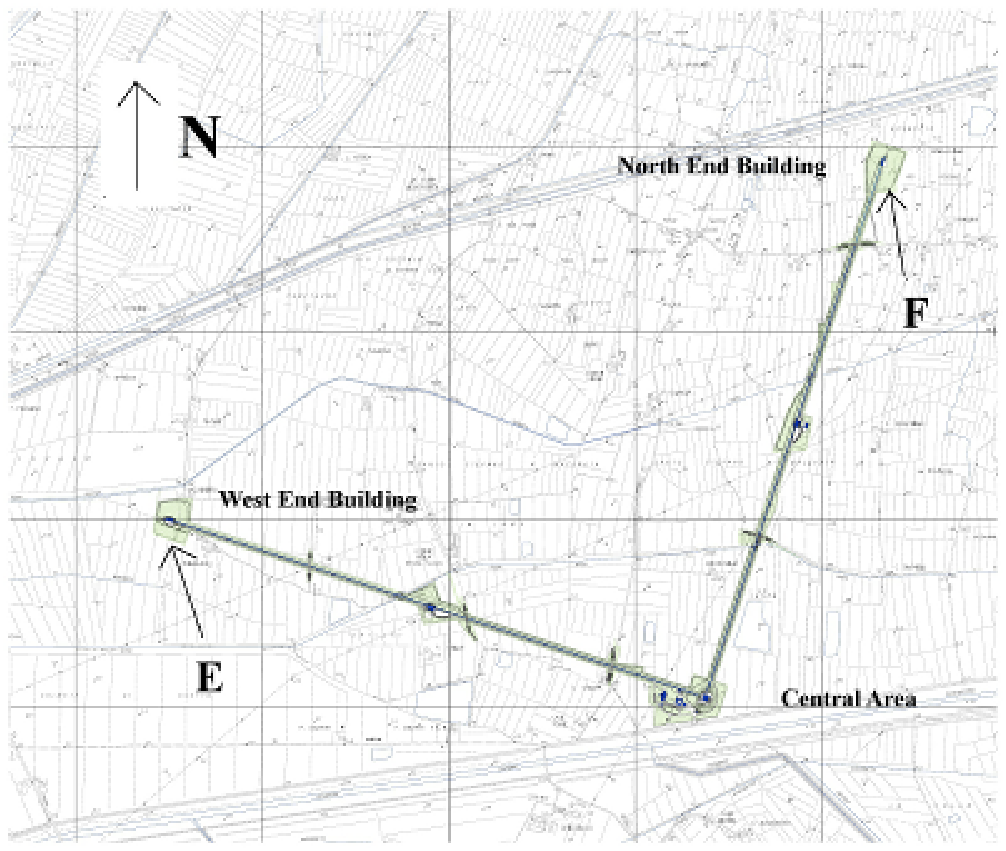}
 \centering \subfigure{}
 \includegraphics[width=3in]{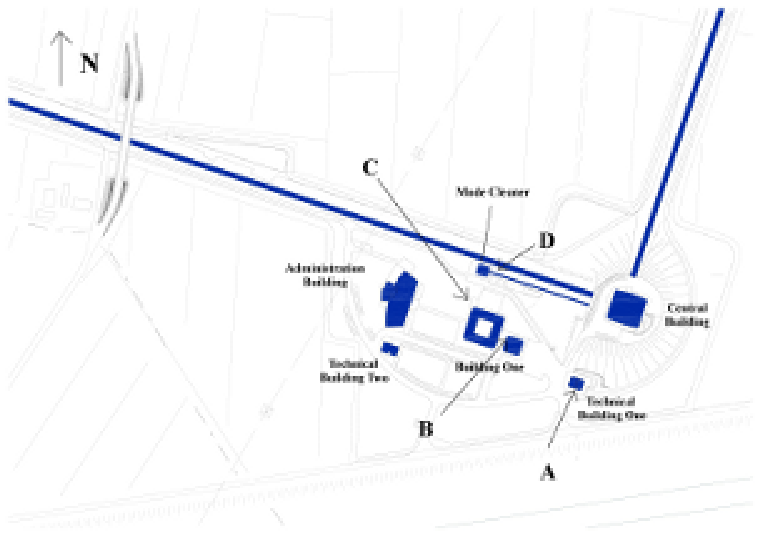}
 \caption{Left: Aerial View of the Virgo Interferometer. Right: Aerial view of Virgo's Central Area. Measurement sites are marked with letters (see Table~\ref{fig:SeismometerLocation}).}
 \label{fig:Virgo}
\end{figure*}

We acquired several sets of data with a Guralp 3TD tri-axial seismometer \cite{Guralp} at six locations around the site. The chosen sampling rate was 250 Hz. The Guralp seismometer is equipped with a GPS antenna receiver to synchronize its data to Virgo seismic channels. Maps for the site and the locations of the measurements can be seen in Fig.~\ref{fig:Virgo}. These locations were chosen for their proximity to machinery known to create large seismic noise. These locations are detailed in Table~\ref{fig:SeismometerLocation}. In each location, the Guralp seismometer was installed on the same concrete platform where the machinery sat and if possible, at a distance of at least a few meters from the machinery to prevent saturation. For each location, the Power Spectral Density (PSD) of the Guralp seismometer, referred to as the ``test probe,'' is computed (see the left of Fig.~\ref{fig:Noisy}). Each measurement lasts approximately 24 hours in order to determine the hourly and daily cycle operation of the machines. This measurement is compared with a tri-axial seismometer, referred to as the ``reference probe,'' permanently stationed in the closest nearby experimental area containing sensitive interferometer components \cite{VirgoEnvironment}. In the Mode Cleaner, West End, and North End Buildings, the reference probe was an Episensor FBA ES-T, while in the Central Building, the probe was a Guralp CNG-T40. If the source is not obvious, suspected machines are analyzed with a piezoelectric accelerometer (PCB) \cite{PCB} placed in direct contact with the machine and read out with a spectrum analyzer. The machines' characteristic frequencies are then compared to those seen in the coherence.

\begin{table}
  \begin{center}
  \scalebox{0.6}{
  \begin{tabular}{|c||c|c|c|c|c|}
  \hline
  \small {Section Letter}& \small {Test Probe Location}& \small{Reference Probe Location}& \small{Distance Between Seismometers (m)} \\
  \hline
  \hline
  \small{A}& \small{Technical Building One: Ground Floor}& \small{Central Building: Ground Floor}& \small{80} \\
  \small{B}& \small{Building One: First Floor}& \small{Central Building: Ground Floor}& \small{105} \\
  \small{C}& \small{Outside Building One: Ground Floor}& \small{Mode Cleaner: Ground Floor}& \small{30} \\
  \small{D}& \small{Outside Mode Cleaner Building: Ground Floor}& \small{Mode Cleaner: Ground Floor}& \small{10} \\
  \small{E/F}& \small{West/North End Building: Ground Floor}& \small{West/North End Building: External Optical Bench}& \small{20} \\
  \hline
  \end{tabular}
  }
  \end{center}
 \caption{Locations of seismometer placement.}
 \label{fig:SeismometerLocation}
\end{table}


\subsection{Analysis Methods}

We considered relevant the frequency lines (persistent spectral peaks) that appear as significantly coherent (coherence rising out of the background) between the test and reference probes, as the source of those lines is very likely to be the same.. Coherence itself does not allow one to distinguish between the seismic signals generated close to the test probe (i.e. machinery under test) from those produced close to the reference probe (i.e. cooling fans of electronic units). In order to determine when the former is the case, the ratio of the test and reference seismometers PSDs is examined, and coherent lines that are stronger near the reference probe are selected. These techniques work well for machines producing continuous lines, such as water pumps and cooling fans.

Periodic lines, on the other hand, are more difficult to identify. Because the lines come and go, they tend to be washed out in PSD averaging. For this reason, frequency-time (FT) plots of the PSDs are produced and examined by eye for periodic lines. In order to identify the source of these lines, we compute the root-mean-square (RMS) around that line, which is defined as:

\begin{equation}
RMS = \sqrt{\sum_{i=f_1}^{f_2}{x_i}^2*{\delta f}}
\end{equation}

where $x_i$ is the ith component of the frequency band in question and $\delta $f is the width of the frequency bins in the spectrum. The sum goes from $i=f_1$ to $i=f_2$, where $f_1$ and $f_2$ are the minimum and maximum frequencies respectively in the frequency band. This RMS value is then correlated to the time series of various Infrastructure Machine Monitoring System (IMMS) signals, including temperature and pressure probes.

\section{Characterization Examples}

In this section, the most significant examples utilizing methods discussed above are presented. In Table~\ref{fig:NoiseLines}, information about important, identified noise lines are given.

\begin{table}
  \begin{center}
  \scalebox{0.6}{
  \begin{tabular}{|c||c|c|c|c|c|}
  \hline
  \small {Location}& \small {Frequency (Hz)}& \small {Periodicity (Period)}& \small {Source}& \small {PSD Ratio} \tabularnewline
  \hline
  \hline
  \small{Technical Building One}& \small{24.2, 48.4,78.6}& \small{Periodic (42 Minutes)}& \small{Cold Water Chiller 1}& \small{16} \tabularnewline
  \small{Building One}& \small{19.3}& \small{Continuous}& \small{Computer Fans}& \small{13} \tabularnewline
  \small{Outside Building One}& \small{48.9}& \small{Continuous (Daytime)}& \small{B1 Chiller}& \small{2} \tabularnewline
  \small{Mode Cleaner}& \small{48.9, 97.8}& \small{Periodic (21.5 Minutes)}& \small{Mode Cleaner Chiller}& \small{220} \tabularnewline
  \small{West End Building}& \small{47.1}& \small{Continuous}& \small{Warm Water Pump}& \small{1.1} \tabularnewline
  \small{West End Building}& \small{48.8}& \small{Continuous}& \small{Cold Water Pump}& \small{6.1} \tabularnewline
  \small{North End Building}& \small{22.8}& \small{Continuous}& \small{Water Pump}& \small{4.2} \tabularnewline
  \hline
  \end{tabular}
  }
  \end{center}
  \caption{Important noise lines identified in the study. The table provides the line's frequency, periodicity, source, and the ratio of the PSDs of the test and reference probes.}
  \label{fig:NoiseLines}
\end{table}

\subsection{Characterization of Noise from Technical Building One}

\begin{figure*}[hbtp!]
 \centering \subfigure{}
 \includegraphics[width=3in]{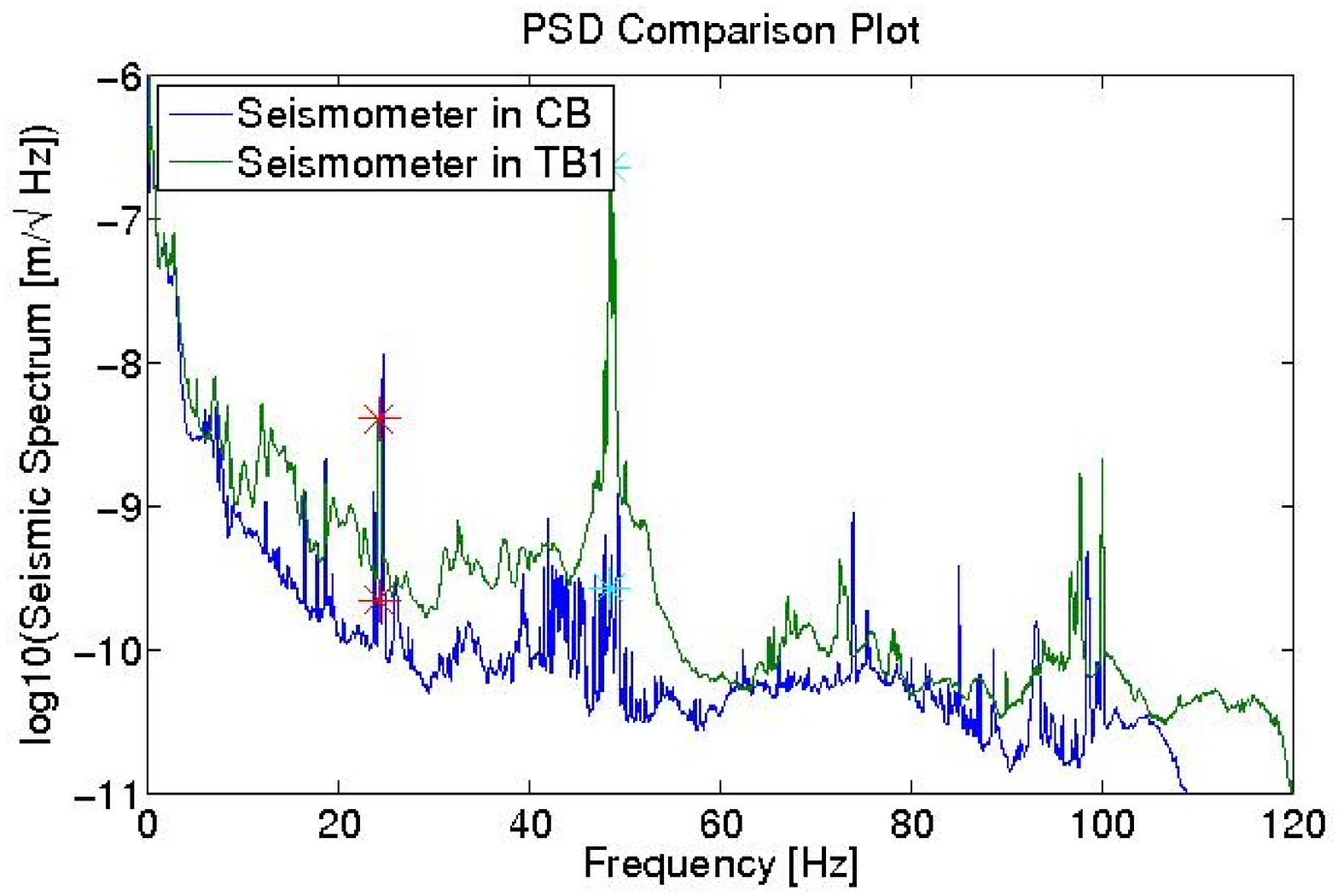}
 \centering \subfigure{}
 \includegraphics[width=3in]{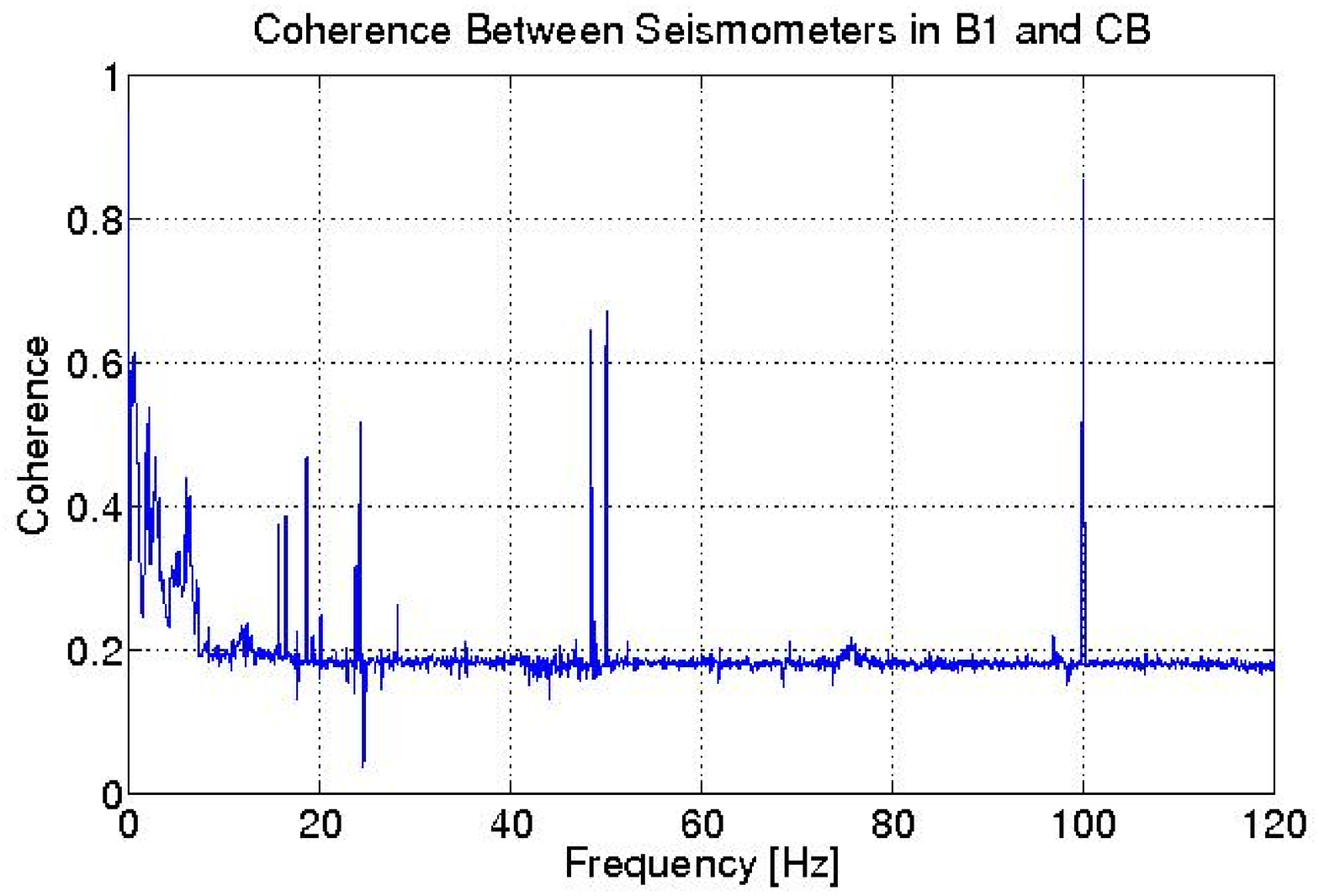}
 \caption{Left: PSDs of the seismometers in TB1 and the CB. The peaks topped by red stars correspond to the 24.2Hz periodic line while the blue stars correspond to the 48Hz continuous line. Right: Coherence between seismometers in TB1 and the CB. The peak topped by the red star corresponds to the 24.2Hz periodic line while the blue star corresponds to the 47.9Hz continuous line.}
 \label{fig:ComparisonTB1CB}
\end{figure*}

\begin{figure*}[hbtp!]
 \centering \subfigure{}
 \includegraphics[width=3in]{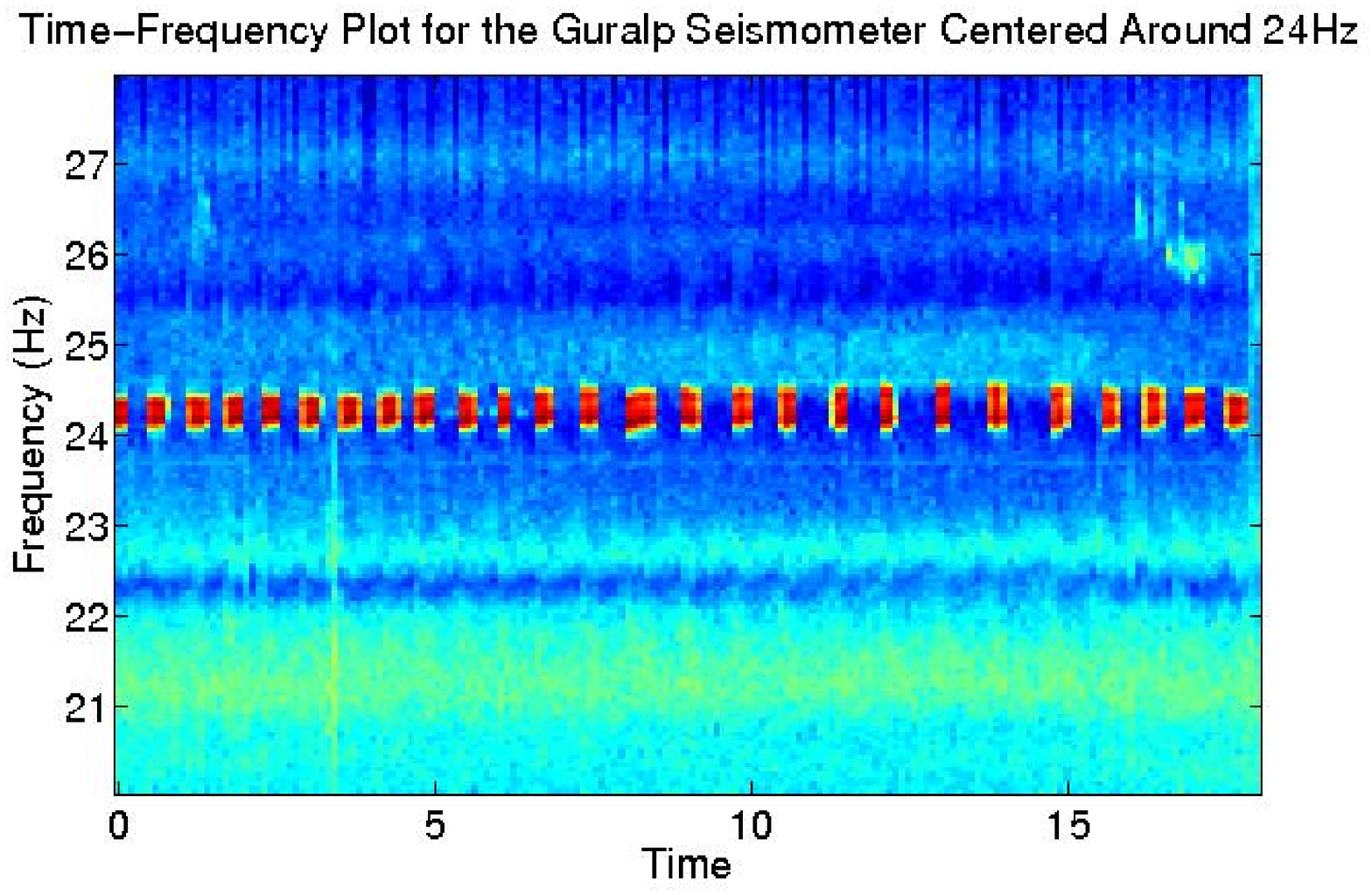}
 \centering \subfigure{}
 \includegraphics[width=3in]{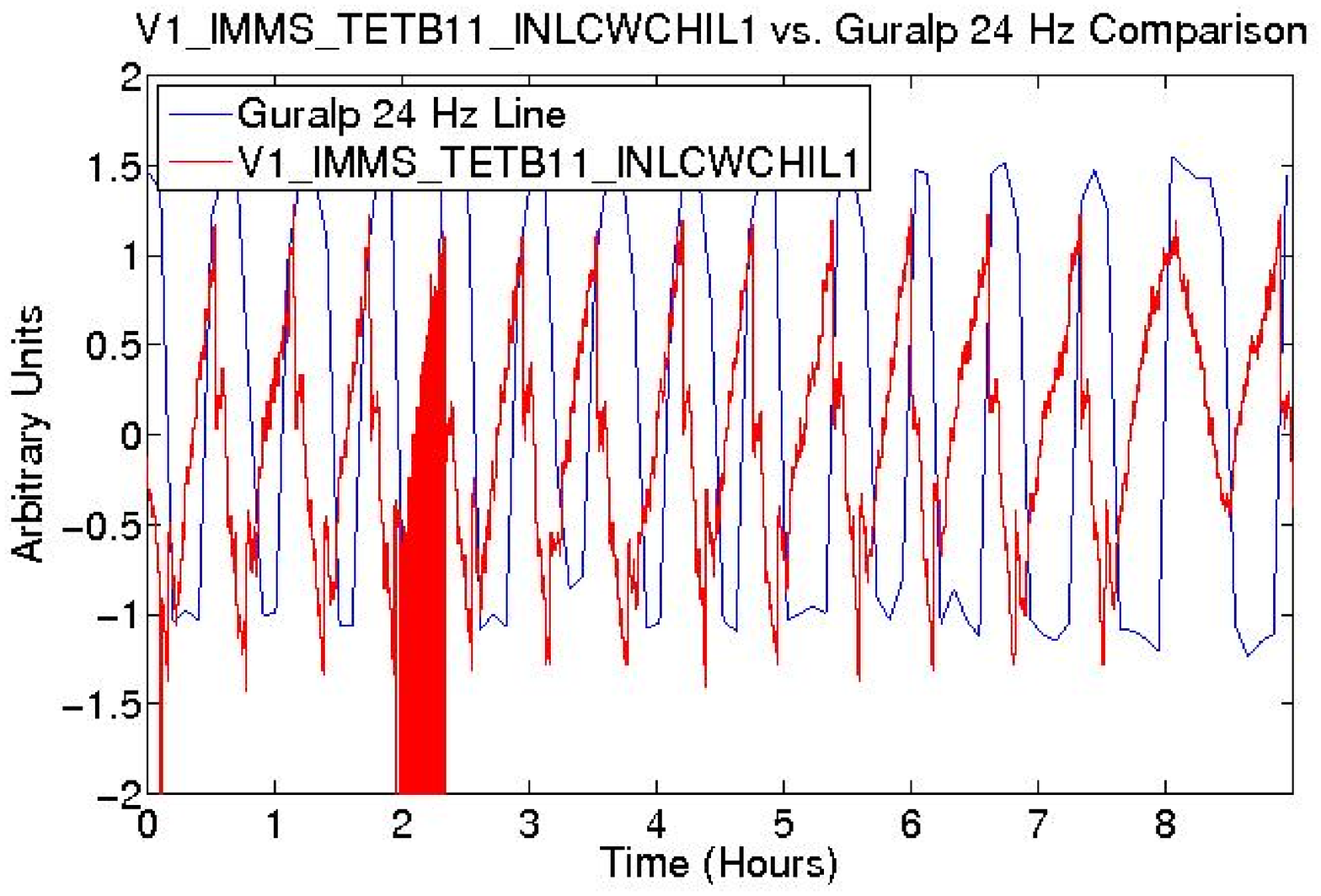}
 \caption{Left: FT-plot of the test probe in TB1 zoomed in on the 24 Hz region. Right: A plot of both the RMS in the 24Hz band of the Guralp seismometer as well as the time series of a temperature monitor of the first (of two) cold water chillers.}
 \label{fig:TB124}
\end{figure*}

A data set was taken in Technical Building One (TB1) and compared with data from a seismometer on the floor of the Central Building (CB), where the two arms of the interferometer converge. These two probes are approximately 80 meters apart. The PSDs of the test and reference probes, as well as the coherence between them, can be seen in Fig.~\ref{fig:ComparisonTB1CB}. A periodic line around 24.2Hz (with harmonics at 48.4Hz and 72.6Hz) is seen, which is showed in detail on the left plot of Fig.~\ref{fig:TB124}.

The coherence for this line between the test and reference probes was about 0.6, indicating a relevant correlation between the two seismometers at that frequency. In order to find the source of the periodic line, it was necessary to determine in which building the line was louder. The ratio of the PSDs (Seismometer in TB1/Seismometer in CB) was approximately 16, indicating the source of the line to be in TB1.

The time series of several machinery monitors in TB1 were compared with the RMS of the 24.2Hz signal in the Guralp seismometer, one of which may be seen on the right of Fig.~\ref{fig:TB124}. From this figure, the RMS is clearly correlated with the time series of a temperature monitor of the first water chiller. We see that when the temperature of the water reaches a high point, the cold water chiller switches on, causing the temperature of the water to decrease. When the temperature reaches a low point, the water chiller switches off and the whole process starts again. The chiller is located on the roof of TB1. Even though the chiller is equipped with insulating springs, the vibrations can probably travel through the rigid water pipes between TB1 and the CB or in the water itself.

\subsection{Characterization of Noise from the Cold Water Chiller Outside the Mode Cleaner}
Another data set was taken outside the Mode Cleaner Building (MCB), on the same concrete platform as the MC's cold water chiller, and compared with the seismometer on the MCB floor. The MCB contains the end mirror of the Input Mode Cleaner, an optical cavity that filters jitter and power noise as well as higher order modes from the beam. The distance between these probes is approximately 10 meters.

\begin{figure*}[hbtp!]
 \centering
 \includegraphics[width=3in]{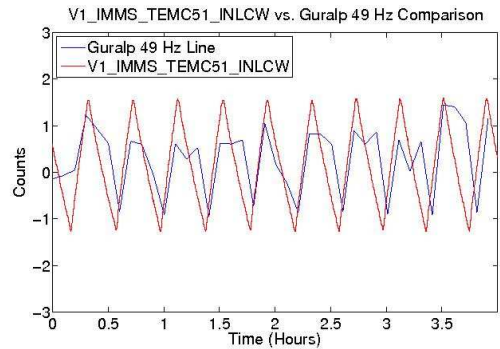}
 \caption{Comparison between the RMS in the 48.9Hz band of the Guralp seismometer and the time series of a temperature monitor of the water chiller.}
 \label{fig:MCChiller}
\end{figure*}

In viewing the PSD plots, a number of significant lines are noticed. This includes continuous lines at 24.2Hz and 38.8Hz and a strong periodic line at 48.9Hz, which has a harmonic at 97.8Hz. To study this line, the RMS around the 48.9Hz band was computed and compared to probes inside the MCB. A water temperature probe follows this line well, as can be seen in Fig.~\ref{fig:MCChiller}, indicating its source is the MC water chiller. This line is important since coherence was noted at its frequency between the GW channel of VSR2 run data and the seismometer in the MCB (see right of Fig.~\ref{fig:Noisy}). Thus, it seems that the MCB water chiller affects the Virgo sensitivity and for this reason, requires prompt attention for noise reduction.

\subsection{Characterization of Noise in the West End Building}
Another data set was taken inside the West End Building (WEB) near the machinery and compared with an Episensor seismometer located on the external optical bench. The distance between these seismic probes is approximately 20 meters.

\begin{figure*}[hbtp!]
 \centering
 \includegraphics[width=4in]{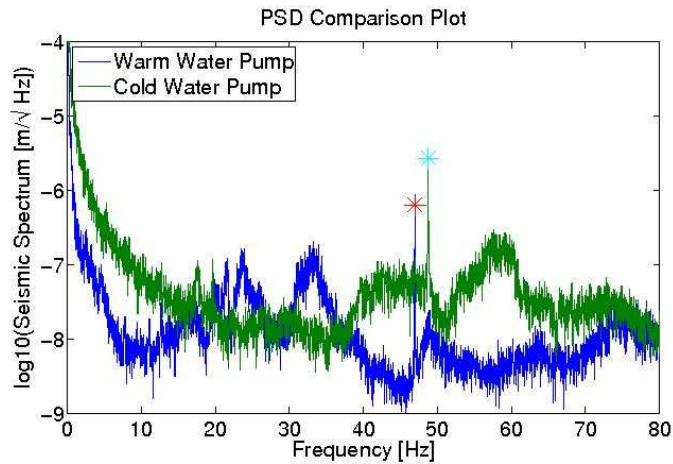}
 \caption{This plot shows the PSD for both the warm and cold water pumps in the WEB. The data were taken with the PCB probe in direct contact with the machines under study. The warm water pump's PSD has a strong line around 47.1Hz, which is shown by the red star on the left. The cold water pump's PSD exhibits a strong line around 48.8Hz, which is shown by the blue star on the right.}
 \label{fig:PCBWE}
\end{figure*}

Analyzing the PSD plots, we noticed several significant lines, including continuous lines at 24.7Hz, 47.1Hz, and 48.8Hz. The 47.1Hz and 48.8Hz are associated with the ``warm'' and ``cold'' water pumps respectively as verified with the PCB probe in Fig.~\ref{fig:PCBWE}. As the 47.1 Hz signal reached the optical bench with almost no attenuation, there is likely a preferred path, for example through water pipes, which must be investigated further.

\section{Attenuation Measurement}

\begin{figure*}[hbtp!]
 \centering
 \includegraphics[width=4in]{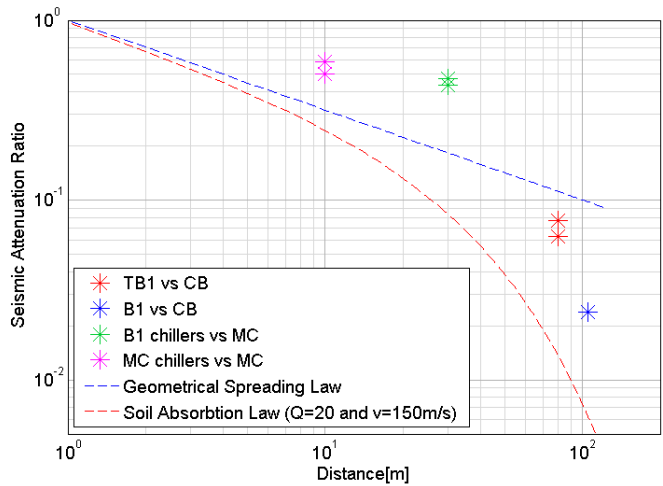}
 \caption{A plot of the average of the PSD ratio of the coherent lines as a function of distance between the probes. We selected locations separated by at least a few wavelengths of soil. The attenuation factors data at 105m are lower limits only.}
 \label{fig:PSDRatioDist}
\end{figure*}

Using the results above, we seek to measure how much noise attenuation occurs when a machine is on its own foundation and a certain distance away from an experimental area. The PSD ratio of test and reference probes is used as a tentative estimate of the attenuation of the vibration, the result for which can be seen in Fig.~\ref{fig:PSDRatioDist}. Measurements are compared to a dissipation law accounting for geometrical spreading of surface cylindrical waves and for energy dissipation in soil \cite{Soil}.

As can be seen in the plot, distances of about 30 meters or less have an attenuation of approximately a factor of 3, while distances of about 80 meters and greater have attenuation values greater than 10. Indeed, the simple noise attenuation model does not apply in most of the examined cases. One reason is that the point-like source approximation  is not valid for short propagation distances (d $\le\lambda$, with our seismic waves length $\simeq$ 10m) or for extended sources (L $\ge$ d, with the machine platform size L $\simeq$ 10m). In addition, the soil medium is often not uniform, and waves' reflection and refraction can occur at discontinuities. Wave amplification due to mechanical resonances of the two platforms (technical area and experimental area) which are excited by the seismic noise is also possible. Another important reason is that in most cases, propagation does not occur uniquely through soil. Less dissipative paths can exist, such as through water pipes or pressure waves inside the water itself.

\section{Conclusion}

Noise from several Virgo infrastructure devices, such as water chillers, heaters, and pumps, seismically affects sensitive parts of the interferometer. Although an impact on the present interferometer is not evident in Virgo's sensitivity (with the exception of the MC chiller), the noise reaching the experimental area is, in several cases, considerably above the background, and requires a reduction for AdV. 

Measured attenuation of seismic signal from a distant source seems not compatible with the much stronger attenuation expected from soil dissipation. It is suspected that water pipes might function as seismic shortcuts and thus need more comprehensive investigation and mitigation attention. Noise transmitted through soil can be more efficiently reduced using seismic isolation systems (i.e. springs).

Further beneficial studies might include setting up a noise source with known power and characteristic frequencies at various distances from a sensitive experimental area. This would allow a comprehensive study of the noise attenuation as a function of distance, something we were unable to do in this study.

\ack

This project is funded by the NSF through the University of Florida's IREU program. The work has been carried on also thanks to the support coming from Italian Ministero dell'Istruzione, dell'Universita' e della Ricerca through grant PRIN 2007NXMBHP.

\section*{References}
\renewcommand\refname{Bibliography}
\bibliographystyle{iopart-num}
\bibliography{VirgoRef}

\end{document}